\documentclass[apj]{emulateapj}
\usepackage{amssymb}
\usepackage{color}
\begin{document}
\title{PLANETESIMAL ACCRETION IN BINARY SYSTEMS: COULD PLANETS FORM AROUND $\alpha$ CENTAURI B ? }
\author{Ji-Wei Xie$^{1,2}$,  Ji-Lin Zhou$^1$, Jian Ge$^2$}
\affil{$^1$Department of Astronomy, Nanjing University,
Nanjing, Jiangsu, 210093, China} 
\affil{$^2$Department of Astronomy, University of Florida, Gainesville, FL, 32611-2055, USA}
\email{xiejiwei@gmail.com}

\begin{abstract}
Stellar perturbations affect planet-formation in binary systems. Recent studies show that the planet-formation stage of mutual accretion of km-sized planetesimals is most sensitive to binary effects. In this paper, the condition for planetesimal accretion is investigated around $\alpha$ CenB, which is believed to be an ideal candidate for detection of an Earth-like planet in or near its habitable zone(0.5-0.9 AU). A simplified scaling method is developed to estimate the accretion timescale of the planetesimals embedded in a protoplanetary disk. Twenty-four cases with different binary inclinations($i_B$=0, 0.1$^o$ 1.0$^o$, and 10$^o$), gas densities(0.3,1,and 3 times of the Minimum Mass of Solar Nebula, MMSN hereafter), and with and without gas depletion, are simulated. We find: (1) re-phasing of planetesimals orbits is independent of gas depletion in $\alpha$ CenB, and it is significantly reached at $1-2$ AU, leading to accretion-favorable conditions after the first $\sim10^5$ yrs, (2)the planetesimal collision timescale at 1-2 AU is estimated as: $T_{col}^B\sim(1+100i_B)\times10^3$  yrs, where $0<i_B<10^o$, (3)disks with gas densities of 0.3-1.0 MMSN and inclinations of 1$^o$-10$^o$ with respect to the binary orbit, are found to be the favorable conditions in which planetesimals are likely to survive and grow up to planetary embryos,  (4)even for the accretion-favorable conditions, accretion is significantly less efficient as compared to the single-star case, and  the time taken by accretion of  km-sized planetesimals into planetary embryos or cores would be at least several times of $T_{col}^B$, which is probably longer than the timescale of gas depletion in such a close binary system. In other words, our results suggest that formation of Earth-like planets through accretion of km-sized planetesimals is possible in $\alpha$ CenB, while formation of gaseous giant planets is not favorable.

\end{abstract}


\section{INTRODUCTION}
Planet formation in binary systems is an important issue, as a major part of solar-type stars were born in binary or multi-stars systems (Duquennoy \& Mayor 1991; Mathieu et al. 2000; Duch{\^e}ne et al. 2007) and approximately $\sim25\%$ of detected exoplanets are expected to inhabit binary star systems (Desidera \& Barbieri 2007). Most of the current detected planet-bearing binary systems are wide S-types(Eggenberger et al. 2004), meaning the companion star acts as a distant satellite, typically orbiting the inner star-planet system over $\sim$ 100 AU away. Nevertheless, at least three planets$-$ GJ86b(Queloz et al. 2000), $\gamma$ Cephei b (Hatzes et al. 2003), and HD41004b (Zucker et al. 2004)$-$are found in close binary systems with stellar separation of only $\sim 20$ AU. In addition, our closest neighbor$-$ $\alpha$ Centauri AB, of which no planet has yet been detected$-$ is also a close binary system with stellar orbital semimajor axis of 23.4 AU and eccentricity of 0.52. The binary proximity and eccentricity can stir strong gravitational perturbations which may pose a threat to planet formation. To address this issue, many researchers(Barbieri et al 2002; Quintana 2004; Quintana \& Lissauer 2006; Quintana \& Lissauer 2007; Quintana et al. 2007; Guedes et al. 2008) have performed a series of N-Body simulations and found Earth-mass planets can form in such close binary systems. However, all their simulations have made an assumption that the planetary disk was initially made of proto-planet of at least Lunar-mass. If Earth-mass planets can form from Lunar-mass protoplanets, the question of whether Lunar-mass embryos can form and remain in stable orbits is critical towards understanding the likelihood of Earth-like planets forming in binary star systems.

\subsection{Planetesimal Accretion in Binary Systems}
According to the current standard planet-formation model, protoplanets or embryos are formed by accretion of 1-10 km-sized (in radii) planetesimals whose own origin is still a big puzzle to our knowledge (Goldreich \& Ward 1973; Weidenschilling \& Cuzzi 1993; Youdin 2008). The accretion from planetesimals to embryos usually follows a runaway mode in single-star systems(Greenberg et al. 1978; Wetherill \& Stewart 1989; Barge \& Pellat 1993; Kokubo \& Ida 1996, 1998, 2000; Rafikov 2003, 2004).  However, things are entirely different for the close binary systems.  

Accretion in binary systems may probably not follow the normal runaway mode or even be suppressed because the binary perturbations stir up the relative velocity ($\triangle V$) among the planetesimals. Recent studies have found that this issue is much more complicated if the binary perturbations are coupled with gas drag forced by the gas disk. Marzari \& Scholl (2000) found that the $\triangle V$ is significantly reduced by the ``phasing" of planetesimals' orbits (``orbital phasing" hereafter) under the coupling of binary perturbations and gas drag. However, Th{\'e}bault et al. (2006, 2008, 2009) have pointed out that the ``orbital phasing" is size-dependent. If assuming a population of planetesimals with a size distribution, such as a Gaussian distribution with a medium of 5 km and range of 1-10 km, the average $\triangle V$ is still too high for accretion in most of the region of binary systems with separations of $\sim$ 20 AU. Paardekooper et al. (2008) investigated this issue with an evolving-disk model and found the conditions for planetesimal accretion were more hostile than in the non-evolving symmetrical disk. To rescue the planet formation in such close binary systems, Xie \& Zhou (2008, 2009) have found two individual mechanisms which may reduce the effects of ``size dependence of orbital phasing" and favor the planetesimal accretion. One mechanism is ``orbital re-phasing" induced by gas depletion. Taking the $\gamma$ Cephei system as an example,  Xie \& Zhou (2008) have found that ``orbital phasing" is significantly restored for all size planetesimals after several timescales of gas depletion. This ``orbital re-phasing" reduces $\triangle V$ to very low values, favoring planetesimal accretion. Another mechanism is small binary inclination. Taking $\alpha$ CenA as an example, Xie \& Zhou (2009) found that the $\triangle V$ can be significantly reduced if a small inclination of such as $1^o-5^o$ is considered between the binary orbit and the gas disk. However, the efficiency of these mechanisms may be quite diverse in different binary systems with different stellar mass ratios, eccentricities, inclinations, and separations. For an example, ``orbital re-phasing" mainly depends on the timescale of gas depletion in $\gamma$ Cephei systems as shown in Xie \& Zhou (2008), while it is found to be nearly independent of gas depletion in $\alpha$ CenB as shown later in this paper. In addition, for highly inclined binary systems($i_B>10^o$), Marzari et al. (2009a) found gas drag has little effect on planetesimals because of their significant vertical excursion above the gas disk plane, and the Kozai mechanism strongly inhibits planetesimal accretion if $i_B$ larger than $40^o$.

A crucial problem that has not been addressed in all previous studies is:  what is the timescale for planetesimal accretion in binary systems.  This timescale can be roughly treated as the collisional timescale $T_{col}$ if most collisions lead to accretion rather than erosion.
In single star systems, $T_{col}$ can be estimated as
$$
T_{col}^{S}={1\over n\triangle V\pi R_p^2}={2a<i>\over\triangle V}{m\over\Sigma_d\pi R_p^2},
$$
\begin{equation}
=4\times10^5f_d^{-1} f_{ice}^{-1}({a\over \rm AU})^{5/2}({\triangle V\over \rm1ms^{-1}})^{-1}({<i>\over 10^{-3}})({R_p\over \rm km})  \ \ \  \rm yrs,
\end{equation}
where $n, m, R_p, a, <i>$ and $\Sigma_d$ are the number density, mass, radii, semi-major axis, average orbital inclination and surface density of planetesimals in the disk. $f_d$ and $f_{ice}$ are scaling numbers of solid surface density and solid enhancement beyond ice line, respectively. Here, $f_d=1$ corresponds to MMSN, i.e., $\Sigma_d=10$gcm$^{-3}$ at 1 AU, and $f_{ice}=4.2$ accounts for solid enhancement beyond the ice line, which usually lies in 2-3 AU for solar type stars. 
Considering the relation, 
$$
\triangle V \simeq \ <e>V_k\simeq 2<i>V_k
$$
\begin{equation}
\simeq60({<i>\over10^{-3}})({a\over \rm AU})^{-1/2}({M_A\over \rm M_\odot})^{1/2}  \ \ \rm ms^{-1} 
\end{equation}
where $V_k$ is the local Keplerian velocity and $<e>$ is the average orbital eccentricity of planetesimals,
Eq(1) can be rewritten as
\begin{equation}
T_{col}^{S}={2\over3}\times10^4 f_d^{-1} f_{ice}^{-1}({a\over \rm AU})^3({M_A\over \rm M_\odot})^{-1/2}({R_p\over \rm km})  \ \ \ \rm yrs.
\end{equation}
This estimation give a typical collisional timescale of $10^4$ yrs roughly for planetesimals with 1-10 km sizes at 1 AU.

In binary systems, however, neither Eq(1) nor Eq(2) is satisfied, and thus the collisional timescale for binary systems ($T_{col}^B$) is not clear. Without any knowledge of $T_{col}^B$, we may misunderstand planetesimal accretion in binary systems. For an example, if $T_{col}^B\sim10^5$ yrs or longer, we may draw wrong conclusions on the conditions of planetesimal accretion by simulations of a binary system for only $10^4$ yrs. The conditions, such as $\triangle V$, in the later several $10^4$ yrs, may become either hostile for accretion by orbital randomization or favorable for accretion by orbital re-phasing.  Therefore, an accurate estimate of $T_{col}^B$ is very crucial to determine whether planetesimal accretion is favored or suppressed in binary systems. In this paper, we develop a simplified method to estimate $T_{col}^B$.

\subsection{Planet Formation in $\alpha$ Centauri B}
$\alpha$ Centauri A and B are main sequence stars with spectral types of G2V and K1V, masses of 1.1 $M_{\odot}$ and 0.93 $M_{\odot}$, respectively. They are bound together as a binary system (or a triple star system if including the M dwarf Proxima Centauri, which orbits the AB pair with a semimajor axis over 10,000 AU according to Wertheimer \& Laughlin 2006) with relative orbital semimajor axis of 23.4 AU and eccentricity of 0.52(Pourbaix et al. 2002). As our closest neighbor in space, $\alpha$ Cen AB has been monitored for over one hundred years, and no planet has been identified around either $\alpha$ CenA or $\alpha$ CenB by far. It is concluded by Endl et al. (2001) that the mass upper limit of a planetary companion is 2.5 $M_{Jupiter}$ for $\alpha$ CenA and 3.5 $M_{Jupiter}$ for $\alpha$ CenB at any orbital radius.  In other words, we still cannot rule out presence of lower mass planets, especially an Earth-like planet. In fact, recent simulations(Quintana et al. 2002; Guedes et al. 2008) have shown Earth-like planets could form around $\alpha$ CenB, and it is the best candidate for searching for Earth-like planets.  Nevertheless, all their simulations focus on the final stage of planet formation, and implicitly assume a disk of planetary embryos as their initial condition. Thus the remaining question is whether these embryos can form. Th{\'e}bault et al. (2009)  recently addressed this problem by analyzing the conditions for planetesimal accretion. They conclude planetary embryos formation through planetesimal accretion seems impossible around $\alpha$ CenB, unless the binary separation was wider in its initial stages. However, their conclusions are limited in the \emph{absolutely coplanar} case, where the inclination between the gas disk and binary stellar orbit is exactly equal to zero($i_B=0$)

In this paper, we numerically reinvestigate the conditions for planetesimal accretion around $\alpha$ CenB. The major contributions will focus on: (1)testing the efficiency of gas-depletion-induced ``orbital re-phasing" in $\alpha$ CenB, (2)developing a scaling method to estimate the timescale of planetesimal accretion in binary systems, and (3)extending the study of Th{\'e}bault et al (2009) by including the effects of binary inclinations.  In section 2, we present the model and results of our simulations. Based on our results, we discuss planet formation in $\alpha$ CenB in section 3. Finally, in section 4, we summarize our major conclusions.

\section{SIMULATIONS}
\subsection{Models and Setup}
Using the fourth-order Hermite integrator(Kokubo et al. 1998), we follow the evolution of $2\times10^4$ test planetesimals around $\alpha$ CenB, under the coupling of the binary's gravity and gas drag force. Planetesimals with random eccentricities of $0-10^{-4}$, random inclinations of $0-5\times10^{-5}$, and random radii of $1-10$ km, are initially orbiting around $\alpha$ CenB from 0.5 AU to an assumed outer disk-edge of 2.5 AU. This outer disk-edge is consistent with those values derived from Artymowicz \& Lubow (1994), Holman \& Wiegert (1999), and Pichardo et al. (2005). Gas drag is modeled by assuming a 3-dimensional non-evolving axisymmetric gas disk as that in our previous work(Xie \& Zhou 2009). Such a gas-disk model neglects the reaction of gas disk to the binary perturbations, which can further increase planetesimals' $\triangle V$ and inhibit their mutual accretion(Paardekooper et al. 2008). Recently, Marzari et al. (2009b) found gas disk is much less excited by the binary perturbations if the disk self-gravity is considered. Therefore, our axisymmetric gas disk model should be a reliable simplification.   

We track all test planetesimal collisions by setting each body an inflated radius of $5\times10^{-5}(r_{col}/AU)^{1/2}$ AU, where $r_{col}$ is the distance from the collision to $\alpha$ CenB. This choice gives a precision of $\sim$1.5m s$^{-1}$ in every impact velocity estimate. These velocity values are then interpreted in terms of accreting or eroding encounters by comparing them to two threshold velocities ( $V_{low}$ and $V_{high}$, see details in Xie \& Zhou 2009). These two threshold velocities are directly related to the specific critical energy $Q^*$ of the planetesimal. Since $Q^*$ is a broad parameter without being well constrained, we then derived $V_{low}$ and $V_{high}$ from a pessimistic $Q^*$ and an optimistic one, respectively(Th{\'e}bault \& Augereau 2007). For an example, if $\triangle V$ is greater than $V_{high}$, then the collision probably leads to erosion rather than accretion.

In total, 25 simulations(see the Table 1) were performed with the consideration of different initial conditions such as: gas disk density($M_d$), binary's inclination($i_B$), and gas depletion. These include three different gas density cases: low(0.3 MMSN), medium(1.0 MMSN) and high cases (3.0 MMSN), where MMSN denotes the Minimum Mass of Solar Nebula (Hayashi 1981). For the binary's inclination, we study four cases ($i_B$=0, 0.1$^o$, 1$^o$ and 10$^o$) since $i_B$ is probably within 10$^o$ for binary systems closer than 30-40 AU (Hale 1994). The timescale of gas depletion is the same as that used in Xie \& Zhou (2008) as $M_d \propto(t/\tau_{dep})^{-1.5}$ with $\tau_{dep}=10^5$ yr.
The  ``run 00" is simulated for a singe-star system, from which the impact rate is normalized to in the other 24 binary cases. The evolving timescale of  ``run 00" is $10^4$ yrs, which is long enough to find enough impacts for statistics. For the remaining binary cases, the evolving timescale is set to $2\times10^5$ yrs, since $T_{col}^B$ is expected to be much longer in inclined binary cases than $T_{col}^S$ . 

\subsection{Estimate of $T_{col}^B$: Scaling Method}
Although the collisional timescale, $T_{col}^B$, in binary systems cannot be derived  as Eq (1) and (3) as in single star systems, we can estimate $T_{col}^B$ if we know the scaling factor between it and $T_{col}^S$. In this paper, subscripts ``S" and ``B", denote the properties of single star systems and  binary systems, respectively.  

Generally, collisional timescale is inversely proportional to the total collision cross-section of the system, namely, 
$$
T_{col}^S= C_S(NR_P^2)^{-1},
$$
\begin{equation}
T_{col}^B= C_B(NR_P^2)^{-1} 
\end{equation}
where $N$ and $R_P$ are the number and typical radii of planetesimals,  and $C_S$ and $C_B$ are the coefficients for single-star systems and binary systems, respectively.
In a real system, $N$ is too large to calculate,  thus we usually take a system with less(such as $n=10^3-10^4$) particles but larger radii(the inflated radius $r=10^{-5}-10^{-4}$ AU in typical cases) for simulations. It has been shown (Brahic, 1976) that a change in the number $N$ of particles or in their radius $R_P$ affects only the collisional rate of the system, which is proportional to the total collision cross-section. 
Therefore, the collisional timescales of simulations $t_{col}$(namely the reverse of impact rate $n_{imp}$) have a scaling relation with the collisional timescales of real systems, such as:
$$
n_{imp}^S=1/t_{col}^S=1/(T_{col}^SNR_P^2/nr^2)=C_S^{-1}nr^2, 
$$
\begin{equation}
n_{imp}^B=1/t_{col}^B=1/(T_{col}^BNR_P^2/nr^2)=C_B^{-1}nr^2.
\end{equation}
Combining Eq(4) and Eq(5), the scaling factor between $T_{col}^S$ and $T_{col}^B$ is written as: 
\begin{equation}
T_{col}^B=T_{col}^S*(n_{imp}^B/n_{imp}^S)^{-1},
\end{equation}
where $n_{imp}^B/n_{imp}^S$ is called ``normalized impact rate".
Since $n_{imp}^B, n_{imp}^S$ can be directly read out from our simulations (runs 0-24), we therefore use Eq(3) and Eq(6) to estimate $T_{col}^B$. 

\subsection{Results}
\subsubsection{radial migration and distribution}
Figure 1 shows the planetesimal number distributions at three different epochs for the 24 binary cases. At the beginning, planetesimals orbit the central star with a uniform distribution(1000 per 0.1 AU, and 20000 in total) from 0.5 to 2.5 AU. As they are subject to the gas drag force, under which they migrate inward, the distribution then changes. The major features of these distribution changes, as shown in figure 1 are:

1)For $i_B\le$ 1$^o$, their evolutions of planetesimal radial distributions are very similar because their inclinations are too small to induce significant difference in radial migration rate. On the other hand, for $i_B=10^o$ (10$^o$ $\sim0.2$ radian, which is comparable to the binary eccentricity of 0.52), the planetesimal migration rate are highly increased, resulting in a significant mass loss near 1 AU. 

2)Since denser gas leads to stronger gas drag and then faster inward drift, planetesimal inward migrations are much more significant in high-mass gas disks without depletion  than in low-mass gas disks with depletion.

3)In nine cases(6 low-mass gas disk cases with $i_B\le1$$^o$, and 3 medium-mass gas disk cases with depletion and $i_B\le1$$^o$, namely runs 1,7,13, 4,10,16, and 5,11,17), very few planetesimals were lost by inward drift in the habitable zone(0.5-0.9 AU) and outer region (such as 1-2 AU). There remains sufficient material for planet formation. 

\subsubsection{normalized impact rate: $n_{imp}^B/n_{imp}^S$}
Figure 2 shows the normalized impact rate($n_{imp}^B/n_{imp}^S$) as a function of radial distance to the central star($\alpha$ CenB). The entire $2\times10^5$ yrs evolution is divided into 3 phases. As shown in figure 2, the major features are summarized as follows.

Comparing different cases of different binary inclinations($i_B=0, 0.1^o, 1.0^o, 10^o$, from left to right, column by column), we find: 

1) For the coplanar case ($i_B$=0), the impact rate in binary systems is highly increased to as much as 1-2 orders of magnitude of the value in the single star case.  This result has also been observed in figure 3 of Xie \& Zhou (2009) and is explained as a result of the increase in the disk volume density induced by the progressive damping of planetesimals' orbital inclinations.

2)As the binary orbit becomes more and more inclined, the impact rates decrease progressively. And there appears to be a good trend: as $i_B$ is increased by an order of magnitude, the impact rate is reduced correspondingly. For an example, at 1.5 AU, the normalized impact rates in the cases of $i_B=0, 0.1^o, 1.0^o, 10^o$ are, respectively: 
$$
\ n_{imp}^B/n_{imp}^S\sim50, \  \  \  {\rm if}\  \  \ i_B=0.0,
$$
$$
\ n_{imp}^B/n_{imp}^S\sim5.0, \  \  \  {\rm if}\  \  \ i_B=0.1^o,
$$
$$
\ n_{imp}^B/n_{imp}^S\sim0.5, \  \  \  {\rm if}\  \  \ i_B=1.0^o,
$$
\begin{equation}
n_{imp}^B/n_{imp}^S\sim0.05, \  \    {\rm if}\  \  \ i_B=10^o.
\end{equation}

Focusing on the evolution of the impact rates during these 3 episodes($t=0-7\times10^4$ yrs, $t=7-14\times10^4$ yrs, and $t=14-20\times10^4$ yrs ), we find:

3)In cases such as runs 3, 21, 24, impact rates are significantly reduced with time because of the loss of most of their planetesimals caused by the inward drift. 

4)Despite similar mass loss by inward drift in runs 3, 9, 15, impact rates in the later two inclined runs do not drop as much as in the coplanar run 3. Furthermore, in most inclined cases(runs 8,12, 14, 18, for examples), although mass losses are significant, impact rates are not reduced or even increased slightly. 

The reason for result 4) can be found in figure 3, which shows evolution of planetesimal orbital elements ($i, \Omega$) in cases of 14 and 17. At the beginning, planetesimals are excited to orbits with different $i$ and $\Omega$, resulting in reduction in the volume density of the planetesimal disk. Then, their orbits have a progressive re-phasing process due to gas damping. This re-phasing of $i$ and $\Omega$ increases the disk's volume density, and then maintains the impact rate as long as the mass lose by gas drag is not too much. For coplanar binary case with $i_B=0$, there is no ``re-phasing" of planetesimals' $i$ and $\Omega$, thus no maintenance or enhancement in impact rate.  

\subsubsection{accretion ratio}
We define a possible accreting collision if its collision velocity is less than $V_{high}$(see the appendix of Xie \& Zhou, 2009 for details).  Figure 4 shows the fraction of possible accreting collisions(hereafter, accretion ratio) during three episodes for the 24 runs. The major results are summarized as follows.

1)During the first  episode (t=0-7$\times10^4$ yrs), a)for the coplanar cases(runs 1-6), the accreting collisions only weigh $\sim$10-20\% because planetesimals' orbital differential phasing increases their relative velocities(Th{\'e}bault et al. 2009); b)for the small inclined cases(runs 7-18) with $i_B$=0.1$^o$ and 1$^o$, the weights of accreting collisions increase to $\sim$40-50\% because small binary inclination reduces the relative velocities by separating the orbits of bodies with different sizes(Xie \& Zhou 2009); c)for the large inclined cases(runs 19-24) with $i_B=10$$^o$, the accretion weights reduce to $\sim$30\% because $i_B$=10$^o$ is sufficient large to pump up considerable planetesimals' orbital inclinations, which give an extra contribution to increase the relative velocities.

2)For the following two episodes($7\times10^4 < t < 20\times10^4$ yrs), we observe an evident increase in the accretion ratio between 1-2 AU in most cases independent of gas depletion. Most of these increases can be up to 80\%, or even close to 100\% in some cases during the last episode: t=14-20$\times10^4$ yrs. However, for the two exceptions, runs 19 and 22, we have not observed significant increase in accretion ratio. Their gas density is only 0.3 MMSN and the binary orbital inclination is as large as 10$^o$. In such conditions, planetesimals are pumped up to highly inclined orbits with little time to experience a considerable gas drag force. These cases become close to a gas-free case(see some similar cases in Marzari et al 2009a) where conditions usually become accretion-hostile because of the progressive orbital randomization(Th{\'e}bault et al. 2006).


\section{DISCUSSIONS}
\subsection{orbital re-phasing}
As shown in figures 3 and 5, the orbital re-phasing is independent of gas depletion in $\alpha$ CenB, and thus it is different from the one identified by Xie \& Zhou (2008) in $\gamma$ Cephei system. To understand the origins and differences of these two kinds of orbital re-phasing, we should review the dynamics of a planetesimal embedded in the gas disk in binary systems (see details in Mazari \& Scholl 2000, Th{\'e}bault et al. 2006, and Xie \& Zhou 2008). Generally, the planetesimal's orbital elements follow 
damped oscillations. In figure 6, for example, the planetesimal's eccentricity follows an oscillation with a frequency of $\mu$ (see Eq.(4) of Th{\'e}bault et al. 2006) under the secular perturbations of the companion star. Then the timescale of  secular oscillation, $T_e$, can be expressed as
$$
T_e=2\pi/\mu
$$
$$
={4\over3}({a\over \rm AU})^{-3/2}({a_B\over \rm AU})^3(1-e_B^2)^{3/2}({M_A\over \rm M_\odot})^{1/2}({M_B\over \rm M_\odot})^{-1}
$$
\begin{equation}\ \ \ \ \ \ \times \ [1+32(a/a_B)^2(1-e_B^2)^{-3}({M_B\over \rm M_\odot})]^{-1} \ \ \rm yrs.
\end{equation}
Because of the gas drag damping, it is also a damped oscillation with a damping timescale $T_d$, which can be expressed as(see the appendix for details)
\begin{equation}
T_d=12f_g^{-1}({R_p\over \rm km})({a\over \rm AU})^{9/4}({a_B\over \rm AU}){(1-e_B^2)\over e_B}({M_A\over  \rm M_\odot})^{-1/2}      \rm \ \ yrs.
\end{equation}
Here  $a$ and $a_B$ are the semi-major axes of planetesimals and the binary star, $M_A$, $M_B$ are the masses of the primary and companion, respectively, $e_B$ is the orbital eccentricity of the binary, and $f_g$ is the scaling number of gas disk density with respect to MMSN. Note that figure 6 is just the solution based on secular perturbation theory (numerical solution of Eq(15) and Eq(16) in Marzari \& Scholl 2000). Although this solution ignores the effects of higher order and short period perturbations, it basically agrees with the direct N-body calculations as shown in figures 3 and 5.
Besides, as this solution only considers the secular perturbations, its solution curves in figure 6 are much more clear than that in figures 3 and 5, thus it is suitable for us to recognize and understand orbital re-phasing.  
 
As shown in figure 6, oscillation is over-damped in the inner region where gas is denser and thus stronger damping, whereas it is under-damped in the outer region. The critical distance($a_c$) to separate these two damped oscillations can be roughly estimated by equating $T_e$ (ignore the high order terms) and $T_d$, from which we have
$$
a_{c}\sim({1\over9})^{4/15} f_g^{4/15} ({R_p\over \rm km})^{-4/15} ({a_B\over \rm AU})^{8/15}
$$
 \begin{equation}
 [e_B(1-e_B^2)]^{4/15} ({M_A\over  M_B})^{4/15} {\rm \ \ AU}.
\end{equation}
Although Eq(10) cannot give an accurate value of $a_c$, it shows the dependences of $a_c$ on $a_B, e_B, f_g, M_A$, and $M_B$. Generally, under-damped oscillation is more favored in binary systems with high mass ratio($M_B/M_A$) and low density of gas disk. 

A reliable $a_c$ can be derived from figure 6 by measuring the position where the oscillation begins to over-damp, namely where the equilibrium eccentricity begins to depart from the forced eccentricity.  
The vertical dashed lines in figure 6 denote the location of $a_c\simeq1.6$ AU for $\alpha$ CenB and of $a_c\simeq3.6$ AU for $\gamma$ Cephei. Beyond the $a_c$, planetesimals with $R_p\ge1$ km follow under-damped oscillation and eventually converge their eccentricities to $e_f$(forced eccentricity). This orbital re-phasing process only needs progressive damping, and it is independent of gas depletion. Within the $a_c$, planetesimals with $R_p\ge1$ begin to follow over-damped oscillation and force their eccentricities to different equilibriums depending on their radii. In this region, orbital re-phasing will be never reached unless the gas density depletes to some low values. In $\gamma$ Cephei case,  over-damped oscillation dominates the whole planetesimals disk, therefore orbital re-phasing require gas depletion as shown in Xie \& Zhou (2008). On the other hand, in $\alpha$ CenB case, under-damped oscillation is significant beyond 1.5 AU, therefore the orbital re-phasing is independent of gas depletion as seen in figures 3 and 5. It is worthy to point out that the over-damped region in $\gamma$ Cephei would shrink to $a_c\simeq2$ AU if assuming a low-mass gas disk of only 1 MMSN rather than the massive gas disk of 10 MMSN adopted by Xie \& Zhou (2008), which matches very well with the dependence of $a_c$ on $f_g$ as shown in Eq(10). 

\subsection{mass left for planet formation}
Our results(figures 3 and 5) show the efficient orbital re-phasing of all planetesimals are reached near 1.5 AU of $\alpha$ CenB after $\sim10^5$ yrs, leading to a ``possible planetesimal-accretion zone" there. This encouraging result, however, may be challenged by several problems. The first one is, as pointed out by Th{\'e}bault et al. (2009), most planetesimals in the ``possible accretion zone" may be removed as a result of inward drift induced by gas drag if the accretion-favorable episode comes too late. Their results shows no object smaller than $\sim$4 km is left beyond 0.8 AU after $2\times10^5$ yrs. This issue, in fact,  is directly related to the size distribution of the initial planetesimal population, which is poorly constrained by current knowledge on the planetesimal formation. In this paper, with the assumption of a random size distribution from 1 to 10 km, the inward drift problem is less severe or even ignorable in some cases, including runs 1, 4, 5, 7, 10, 11, 13, 16, and 17. Furthermore, the outer edge of planetesimal disk we considered (2.5 AU) is larger than that(1.5AU) of Th{\'e}bault et al.(2009). Therefore, in the possible accretion region, say near 1.5 AU, the mass lost by the inward drift of local planetesimals can be replenished by the planetesimals from outer disk(see the left panel of figure 7). In addition, if the ice line is considered at 2 AU, the supplement of planetesimals is very considerable because of the enhancement of solid mass beyond the snow line(see the right panel of figure 7).  In such a case, gas inward drift does not induce a mass loss problem but a mass enhancement favoring the planetesimal accretion at 1- 2 AU of $\alpha$ CenB. For these considerations, we conclude, mass loss by the inward drift is negligible if $M_d \le 1$ MMSN and  $i_B<10$$^o$.  

\subsection{accretion-hostile transition}  
The second problem is that even when mass is preserved in the possible accretion zone, say 1-2 AU, the accretion-favorable conditions shown in figure 4 are only reached after an accretion-hostile transition duration of at least several $10^4$ yrs. The question that remains, as pointed out by Th{\'e}bault et al. (2009), is how these objects might survive this long erosion period. During this period, it is likely most large planetesimals will be fragmented into small debris which will be quickly removed by the inward drift because of their small sizes. To understand the details of this erosion process one needs a N-body code including fragmentation which is out of the scope of this paper. Nevertheless, we can reach some qualitative conclusions by comparing the collisional timescale($T_{col}^B$) with the duration of the accretion-hostile transition($T_{aht}$). The argument is the following: if the disk is fragmented into small debris, every planetesimal should have experienced at least one erosive collision, or in other words, the duration of the accretion-hostile transition should be longer than the collisional timescale($T_{aht}> T_{col}^B$). On the other hand, if $T_{col}^B> T_{aht}$, then the planetesimals are able to survive the accretion-hostile transition .  As shown in figure 4, $T_{col}^B$ should be at least $10^5$ yrs to survive the accretion-hostile transition.

According to Eq.(1),  the collisional timescale in single star systems is about $T_{col}^S=10^4-10^5$ yrs for 1-10 km-sized planetesimals at 1.5 AU where the center of possible accretion zone is located(see figure 4). If we take a medium value of $T_{col}^S=5\times10^4$ yrs, then adopting the scaling factors of Eq(6) and Eq(7), the collisional timescales at 1.5 AU for different binary cases are roughly estimated as:
$$
T_{col}^B\sim10^3 \ {\rm yrs}, \  \  \  {\rm if} \  \  \ i_B=0.0,
$$
$$
T_{col}^B\sim10^4 \ {\rm yrs},   \  \ \  {\rm if} \  \ i_B=0.1^o,
$$
$$
T_{col}^B\sim10^5 \ {\rm yrs}, \  \  \  {\rm if}  \  \ i_B=1.0^o,
$$
\begin{equation}
T_{col}^B\sim10^6 \ {\rm yrs}, \  \  \  {\rm if}\  \  \ i_B=10^o,
\end{equation}
or expressed as a simple formula as:
\begin{equation}
T_{col}^B\sim(1+ 100i_B)\times10^3  \ \ \ {\rm yrs}, \  \  \ {\rm if}\  \  \ 0\le i_B\le10^o.
\end{equation}

Therefore, for the cases with $i_B > 1.0$$^o$ (runs 13-24), collisional timescale is long enough for planetesimals to survive the accretion-hostile transition. Planetesimals are probably able to accrete into planetary embryos for the cases with $1^o< i_B < 10^o$, while cases with $i_B\ge10^o$ are less favorable for planetesimal accretion, even though their $T_{col}$ are long enough to survive the accretion-hostile phase. The reasons are the following: a)the accretion timescale is so long($T_{col}\ge10^6$ yrs if $i_B\ge10^o$) that almost all the planetesimals would be removed before they have accreted into bigger ones(see cases 20, 21, 23, 24 in figure 1), and b) $i_B\ge10^o$ would lead systems close to gas-free case, in which conditions become accretion-hostile(see cases 19, 22 in figure 4). 

For the coplanar cases($i_B=0$), as shown in figure 4(runs 1-6), more than $80\%$ collisions lead to erosion at the first several $10^4$ yrs. This ``erosion-dominated phase" is longer than the collisional timescale by 1-2 orders of magnitude,  thus planetesimal disk will be efficiently fragmented into debris before conditions to become favorable for planetesimal accretion. This result is consistent with the conclusion of Th{\'e}bault et al. (2009)

For the small inclined cases($i_B=0.1^o$, runs 7-12), during the phase of the first several $10^4$ yrs($1-7\times10^4$ yrs as in figure 4) the erosion collision still weighs a large part ($\sim60\%$), but not as dominant as in the coplanar cases. The chances for accretion and erosion are nearly equal during this period, and thus the final result of the collisional evolution during this phase is uncertain. This ``uncertain phase" dominates the fate (accretion or erosion) of the planetesimal disk, since its duration ($\sim10^5$ yrs) is much longer than the collisional timescale($T_{col}\sim10^4$ yrs, if $i_B=0.1^o$). For this reason, the cases($i_B$=0.1$^o$) are marginal cases, for which we cannot draw any conclusion about whether planetesimals are possible to accrete into planetary embryos or not.

\subsection{planetesimals grow slowly}
As discussed above, planetesimal accretion is possible, most probably at 1-2 AU around $\alpha$ CenB if $1^o<i_B<10^o$. Nevertheless, the accretion process in $\alpha$ CenB will be never like the normal accretion around a single star(type I runaway mode), since most relative velocities among planetesimals are pumped up to values larger than their escape velocities. The accretion may probably follow a type II runaway mode which is characterized with initially a slow orderly growth and then switches to runaway mode whenever the biggest body's escape velocity is larger than $\triangle V$(Kortenkamp et al. 2001; Th{\'e}bault et al. 2006, 2009; Xie \& Zhou 2009). However, even for the type II runaway mode, planetesimal accretion, is significantly slowed down compared to the normal accretion or type I runaway accretion. As shown in figure 2 of Kortenkamp et al. 2001, the planetesimal accretion timescales are $\sim10^5$ yrs and $\sim5\times10^5$ yrs at 1.5 and 2 AU, respectively. Although these results from  Kortenkamp et al. 2001 are based on the Solar-Jupiter system, they have shown a similar mechanism(the type II runaway mode) which can be applied to $\alpha$ CenB.

Besides, accretion process is further slowed down by binary inclination, which stirs the planetesimal disk to a larger scale height with lower volume density, and thus leading to lower impact rate. For the possible accretion favorable cases($i_B=1^o-10^o$) discussed in the last subsection, the collisional timescale is as long as $T_{col}^B\sim10^5-10^6$ yrs, i.e., it takes at least $10^5-10^6$ yrs to form planetary embryos or planetary cores. On the other hand, gas disks in close binary systems may deplete very fast (Cieza et al. 2009). For binary systems such as $\alpha$ CenB with a separation of only $\sim20$ AU, the gas depletion timescale can be as short as a few $10^5$ yrs.
Therefore, in the case of $\alpha$ CenB, although planetesimal accretion is possible, the accretion timescale would be comparable or longer than the gas depletion time scale. In such a case, even if Earth-like planets or planetary cores could be formed, there would be no gas left for them to accrete into gaseous planet. This argument may probably account for the fact that no gas giant has been found orbiting $\alpha$ CenB(Guedes et al. 2008).

\section{CONCLUSIONS}
We perform a series of simulations to investigate the conditions for planetesimal accretion from 0.5-2.5 AU around $\alpha$ CenB. In all of the simulations, we vary the gas-disk density(0.3-3 MMSN), binary inclination(0-10$^o$), and consider both the cases with and without gas depletion. Three indicators$-$planetesimal number(or mass) distribution, planetesimal impact rate(or collisional timescale), and the fraction of accreting collisions$-$ are used in our discussions to determine whether planetesimals can accrete into planetary embryos or planets around $\alpha$ CenB.
Our major conclusions are summarized as following:

1)Most planetesimals are able to remain in the ``possible accretion zone" (1-2 AU around $\alpha$ CenB) against the gas drag induced inward drift if the gas density is relatively small (0.3-1.0 MMSN) and the binary inclination is not too large($i_B<10^o$).

2)For all the cases with and without gas depletion, within a timescale of $\sim10^5$ yrs, we have observed an orbital re-phasing for planetesimals at $\sim1-2$ AU around $\alpha$ CenB(see figures 3 and 5), leading to an accretion favorable condition there(see figure 4). This orbital re-phasing is caused by the under-damped oscillation of planetesimals' eccentricities and inclinations, which is independent of gas depletion and thus different from the orbital re-phasing identified as in $\gamma$ Cephei (Xie \& Zhou 2008). 

3)We develop a simplified scaling method to estimate the collisional timescale in binary systems, $T_{col}^B$. We find $T_{col}^B$ sensitively depends on the binary inclination($i_B$) as $T_{col}^B\sim(1+ 100i_B)\times10^3$  yrs, if $0\le i_B\le10^o$, at $1-2$AU around $\alpha$ CenB. Based on the estimate of $T_{col}^B$, we find, planetesimals are probably able to survive the accretion-hostile transition period in the first $10^5$ yrs and accrete into larger planetary embryos if $1^o<i_B<10^o$.

4)Planetesimal accretion into planetary embryos in $\alpha$ CenB, though possible, take a long timescale as much as $10^5-10^6$ yrs, which is 1-2 orders of magnitude longer than that in normal accretion process in single-star systems. Therefore, the formation of gaseous giant planets, like Jupiter and Saturn, are not favored in $\alpha$ CenB, since gas depletes very fast in close binary systems(Cieza et al. 2009) in a timescale as short as $10^5$ yrs.

In summary, although planetesimal accretion in $\alpha$ CenB is significantly less efficient and slowed-down as compared to single star systems, it is still possible if gas density is 0.3-1.0 MMSN and binary inclination is $1^o<i_B<10^o$. These accretion favorable conditions, in fact, are typical values for initial gas density(Andrews \& Williams 2005) and binary inclination(Hale 1994, Jensen et al. 2004, Monin et al. 2004, 2006) from current observations. Our results support recent work by Guedes et al. (2008), which has shown Earth-mass planets can be formed near the habitable zone(0.5-0.9 AU) of $\alpha$ CenB if the disk is initially composed of lunar-mass planetary-embryos. The possible accretion zone shown in this paper is roughly between 1-2 AU, which matches well with their planet formation zone($\sim$0.5-2.0 AU, as shown in figures 1 and 2 of Guedes et al. 2008). In addition, at the time of writing this paper, we note a promising result from Payne et al. (2009) that Earth-like planes can also form in the habitable zone of $\alpha$ CenB-like binary systems through outward migration from the inner accretion-unperturbered zone(within $\sim 0.7$ AU). Therefore, by combining these studies(Guedes et al. 2008; Payne et al. 2009) and our simulations, it is quite possible that a habitable Earth-like planet may be hidden around $\alpha$ CenB. 

\acknowledgments
We thank Th{\'e}bault, P. for discussions and valuable suggestions.
 This work was supported by the National Natural Science Foundation of China
  (Nos.10833001, 10778603 and 10925313), the National Basic Research Program of China(No.2007CB814800), NSF with grant AST-0705139, NASA with grant NNX07AP14G,
W.M. Keck Foundation and also University of Florida.

\begin{center} 
\bf{APPENDIX}
\end{center}
\appendix
\section{GAS DAMPING TIMESCALE: $T_{d}$}
According to Adachi, Hayashi, \& Nakazawa (1976), the average decay of planetesimal eccentricity caused by gas drag can be given as
\begin{equation}
{1\over e}({de\over dt})=-{1\over \tau_{aero}}({5\over8}e^2+{1\over2}i^2+\eta^2)^{1/2},
\end{equation}
where $\eta\sim0.002(a/AU)^2$, which describe the sub-Keplerian motion of gas, and $e$, $i$ are the orbital eccentricity and inclination of the planetesimal. $\tau_{aero}$ is the characteristic timescale, which is related to the planetesimal size and gas density. With MMSN modeling the gas disk, $\tau_{aero}$ can be typically expressed as
\begin{equation}
\tau_{aero}=15f_g^{-1}({R_p\over km})({a\over AU})^{13/4}({M_A\over M_\odot})^{-1/2}        \ \ \ \rm {yrs},
\end{equation}
where $R_p$ and $a$ are the radius and semi-major axis of the planetesimal, $M_A$ is the mass of primary star, and $f_g$ is the scaling number of gas density with respect to MMSN.
In the binary case of our interest(slightly inclined binary systems), $e$ is much larger than $i$ and $\eta$, thus the time scale for damping $e$ is 
\begin{equation}
T_d=e/{de\over dt}\sim({8\over5})^{1/2}{\tau_{aero}\over e}.
\end{equation}
As long as eccentricity is dominated by secular perturbations, we can take $e$ as an average of forced eccentricity (see Th{\'e}bault et al. 2006) as
\begin{equation}
e_{forced}={5\over4}{a\over a_B}{e_B\over 1-e_B^2},
\end{equation}
then we finally get the damping timescale
\begin{equation}
T_d=12f_g^{-1}({R_p\over km})({a\over AU})^{9/4}({a_B\over AU}){(1-e_B^2)\over e_B}({M_A\over M_\odot})^{-1/2}       \ \ \ \rm {yrs}.
\end{equation}
\\
Note this formula will be invalid if $e_B$ is close to zero, since the evolution is then dominated by others, such as short period perturbations, rather than by secular perturbations.

\clearpage


\begin{table}
  \caption{Initial Conditions for 25 Simulations.\label{tbl-1}}
  \begin{tabular}{@{}lllllllllrrr@{}}
\tableline
\tableline
 & Run  & $i_B$ & $M_d$ & $\tau_{dep}$ \\
 \tableline
 & 00 &  - & 1 & +$\infty$ \\
 & 01 &  0 & 0.3 & +$\infty$\\
 & 02 &  0 & 1 & +$\infty$\\
 & 03 &  0 & 3 & +$\infty$\\
 & 04 &  0 & 0.3 & $10^5$\\
 & 05 &  0 & 1 & $10^5$\\
 & 06 &  0 & 3 & $10^5$\\
 & 07 &  0.1 & 0.3 & +$\infty$\\
 & 08 &  0.1 & 1 & +$\infty$\\
 & 09 &  0.1 & 3 & +$\infty$\\
 & 10 &  0.1 & 0.3 & $10^5$\\
 & 11 &  0.1 & 1 & $10^5$\\
 & 12 &  0.1 & 3 & $10^5$\\
 & 13 &  1 & 0.3 & +$\infty$\\
 & 14 &  1 & 1 & +$\infty$\\
 & 15 &  1 & 3 & +$\infty$\\
 & 16 &  1 & 0.3 & $10^5$\\
 & 17 &  1 & 1 & $10^5$\\
 & 18 &  1 & 3 & $10^5$\\
 & 19 &  10 & 0.3 & +$\infty$\\
 & 20 &  10 & 1 & +$\infty$\\
 & 21 &  10 & 3 & +$\infty$\\
 & 22 &  10 & 0.3 & $10^5$\\
 & 23 &  10 & 1 & $10^5$\\
 & 24 &  10 & 3 & $10^5$\\
\tableline
\end{tabular}
\tablecomments{The units of $i_B$, $M_d$ and $\tau_{dep}$ are $deg$, MMSN, and $yr$ respectively.}
\end{table}



\clearpage
\begin{figure}
\begin{center}
\includegraphics[width=\textwidth]{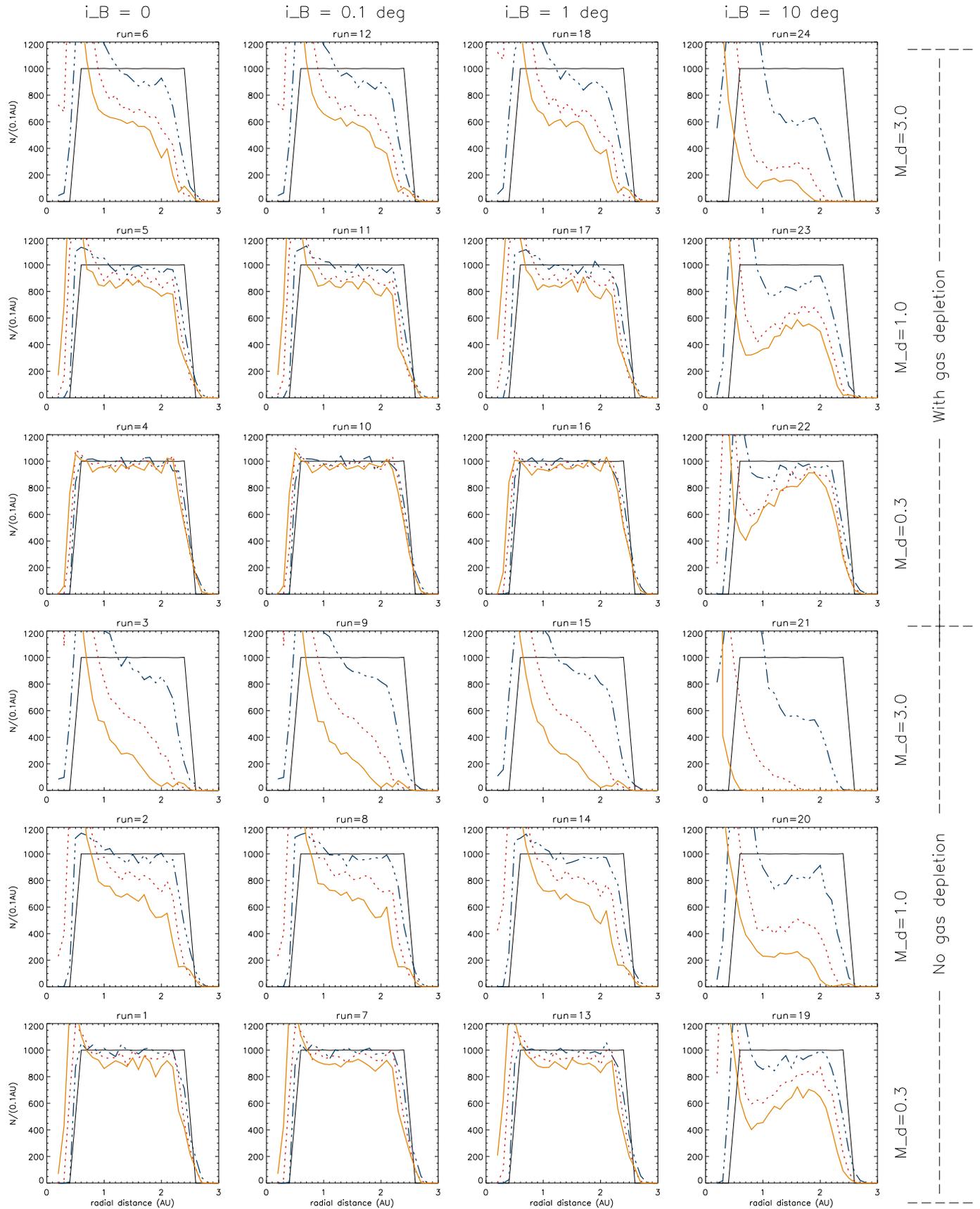}
  \caption{Planetesimal number distribution (N/(0.1 AU)) vs radial distance (AU) to the central star. Trapezoid: t=0, dashed-dot: t=3$\times10^4$ yr, dashed: t=10$\times10^4$ yr, and solid: t=18$\times10^4$ yr }
   \end{center}
\end{figure}

\clearpage
\begin{figure}
\begin{center}
\includegraphics[width=\textwidth]{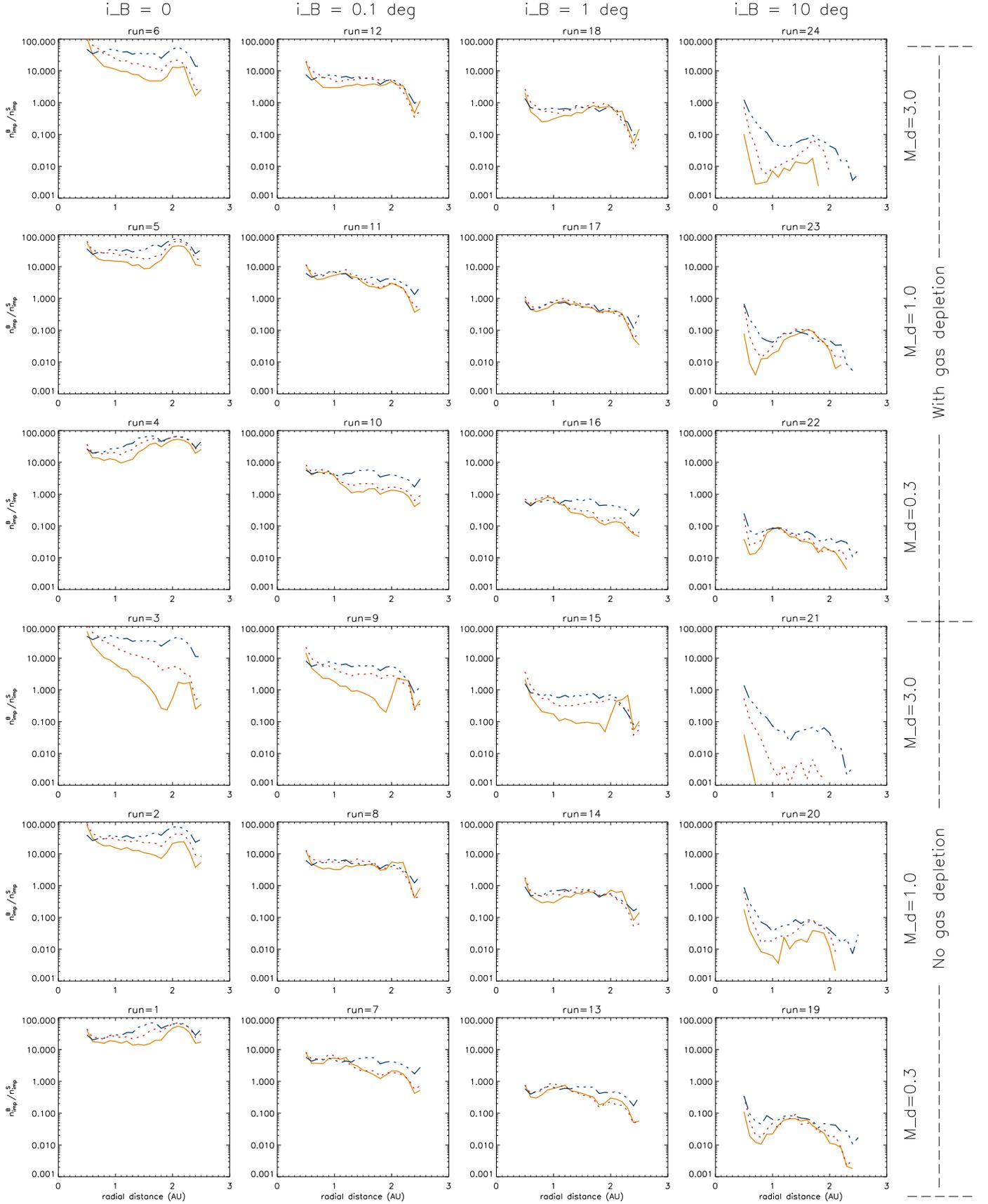}
  \caption{Normalized impact rate($n_{imp}^B/n_{imp}^S$) vs radial distance (AU) to the central star for three phases. Dash-solid: t=0-7$\times10^4$ yr, dot: t=7-14$\times10^4$ yr, and solid: 14-20$\times10^4$ yr.}
   \end{center}
\end{figure}

\clearpage
\begin{figure} 
\begin{center}
\includegraphics[width=\textwidth]{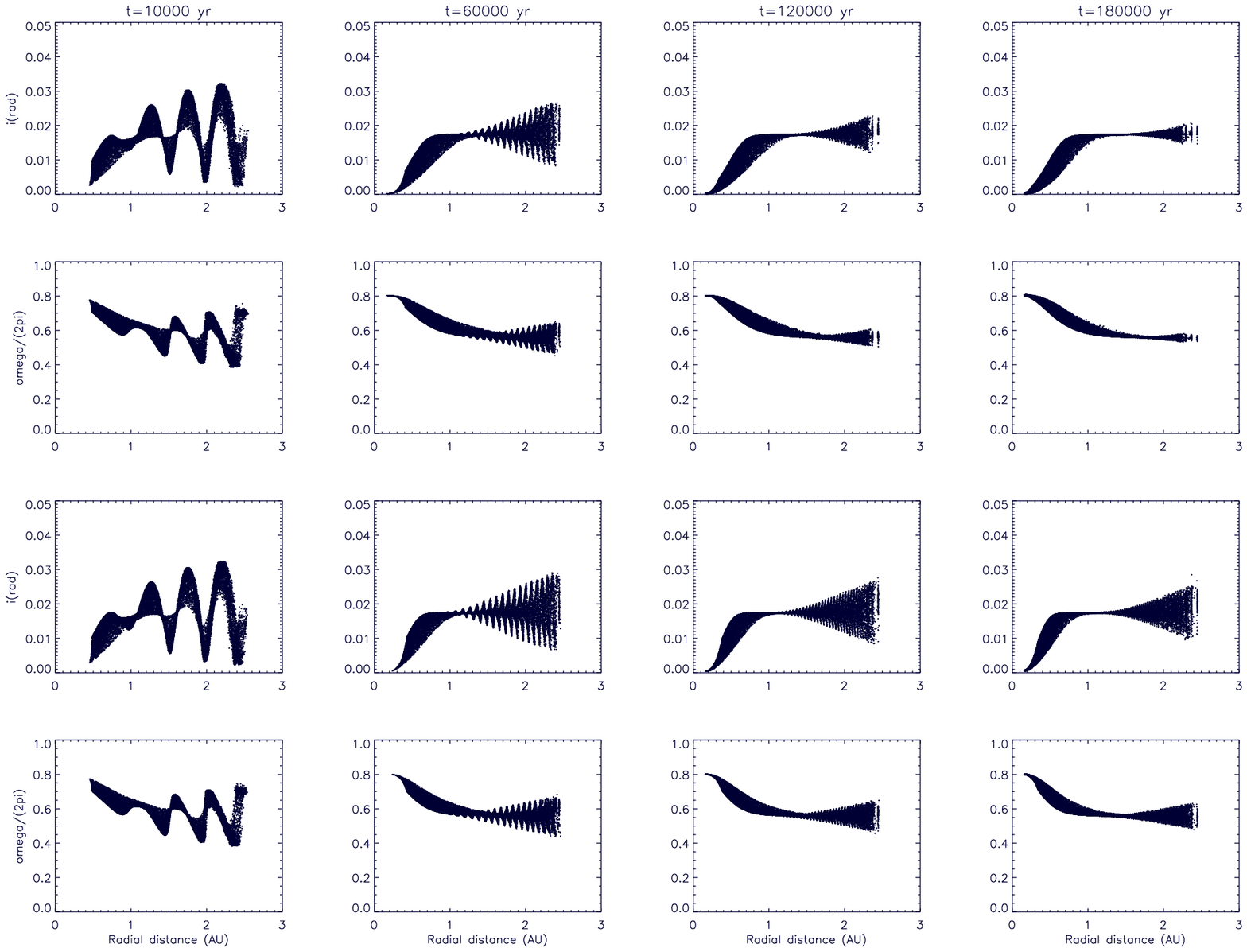}
  \caption{Planetesimal orbital elements (i, $\Omega$) vs radial distance (AU) at 4 different epochs: 1, 6, 12, 18 $\times10^4$ yr for the case 14 with gas depletion(up 2 rows of panels) and case 17 without gas depletion(bottom 2 rows of panels). }
   \end{center}
\end{figure}

\clearpage
\begin{figure}
\begin{center}
\includegraphics[width=\textwidth]{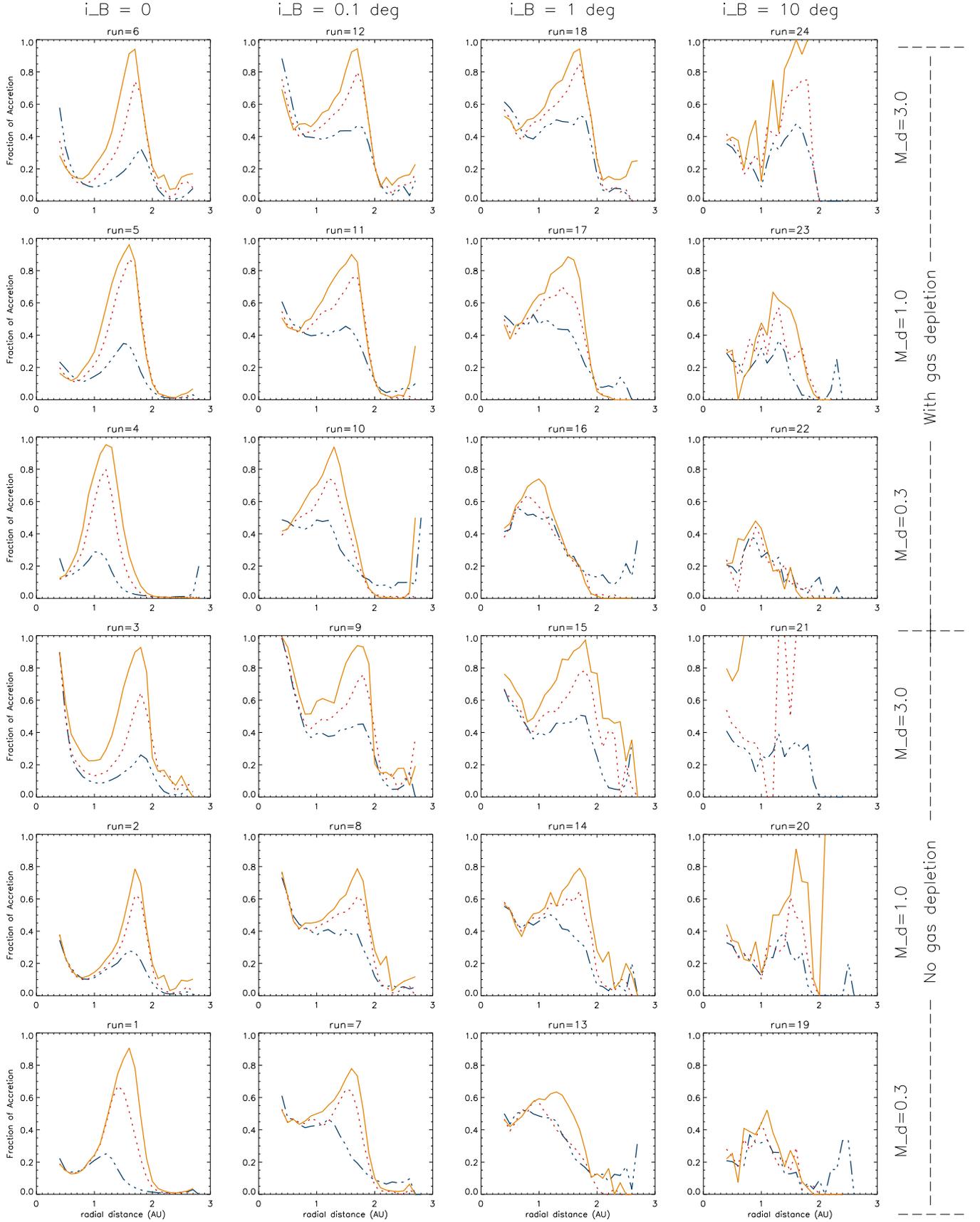}
 \caption{Fraction of possible accreting collisions (normalized) vs radial distance (AU) to the central star. Dash-solid: t=0-7$\times10^4$ yr, dot: t=7-14$\times10^4$ yr, and solid:14-20$\times10^4$ yr.}
\end{center}
\end{figure}

\clearpage
\begin{figure}
\begin{center}
\includegraphics[width=\textwidth]{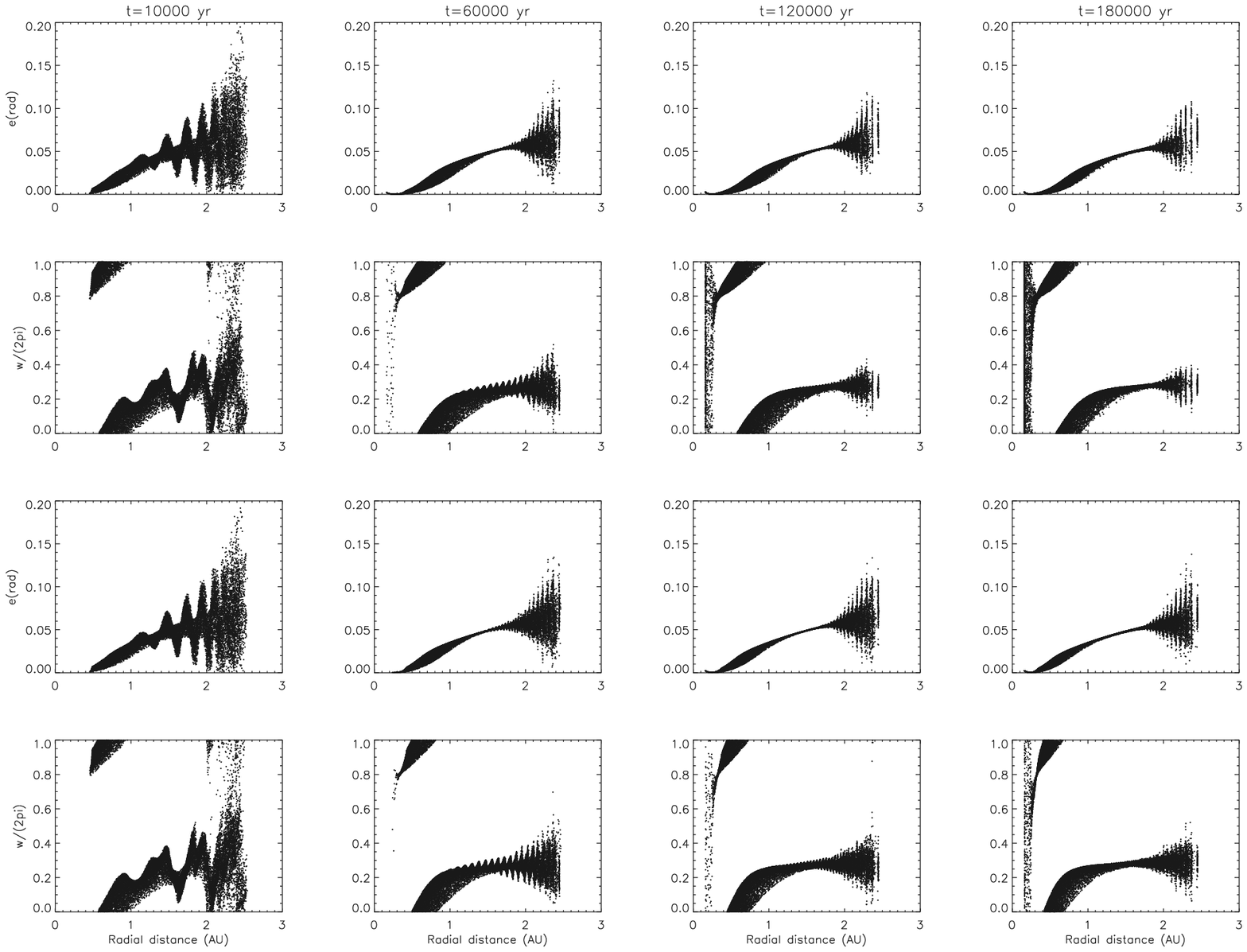}
  \caption{Planetesimal orbital elements (e, $\omega$) vs radial distance (AU) at 4 different epochs: 1, 6, 12, 18 $\times10^4$ yr for the case 14 with gas depletion(up 2 rows of panels) and case 17 without gas depletion(bottom 2 rows of panels). }
   \end{center}
\end{figure}

\clearpage
\begin{figure}
\begin{center}
\includegraphics[width=\textwidth]{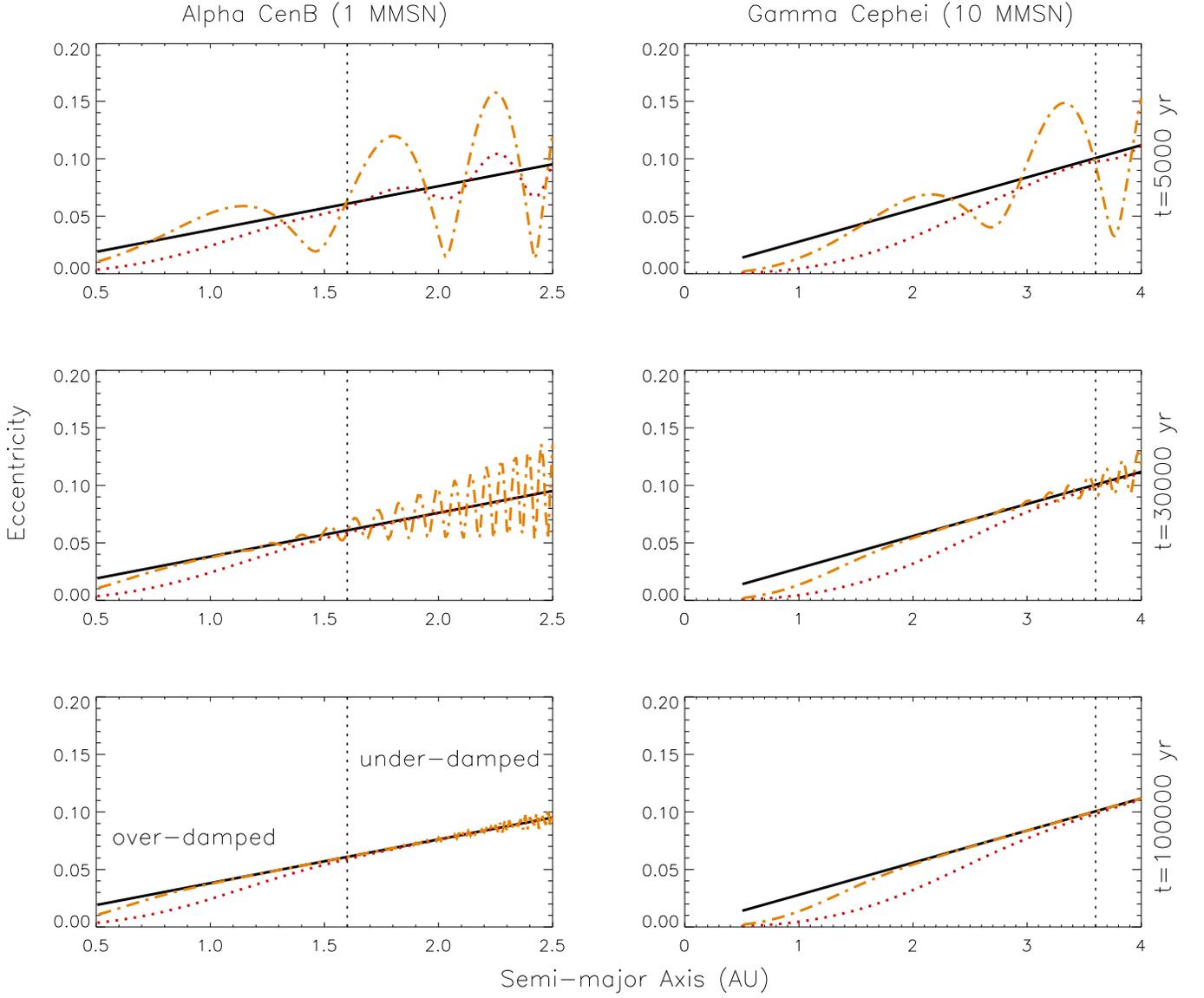}
  \caption{Planetesimal eccentricity vs radial distance (AU) at 3 different epochs(from top to bottom): 5, 30, 100 $\times10^3$ yr for the $\alpha$ CenB with gas density of 1 MMSN(left 3 panels) and for the $\gamma$ Cephei with gas density of 10 MMSN(right 3 panels). Solid: forced eccentricity, dot: eccentricity of a planetesimal with $R_p=1$ km, and dashed-dot: $R_p=10$ km. The vertical dashed line denotes the location of $a_c$, which separate over-damped and under-damped regions.}
   \end{center}
\end{figure}

\clearpage
\begin{figure}
\begin{center}
\includegraphics[width=\textwidth]{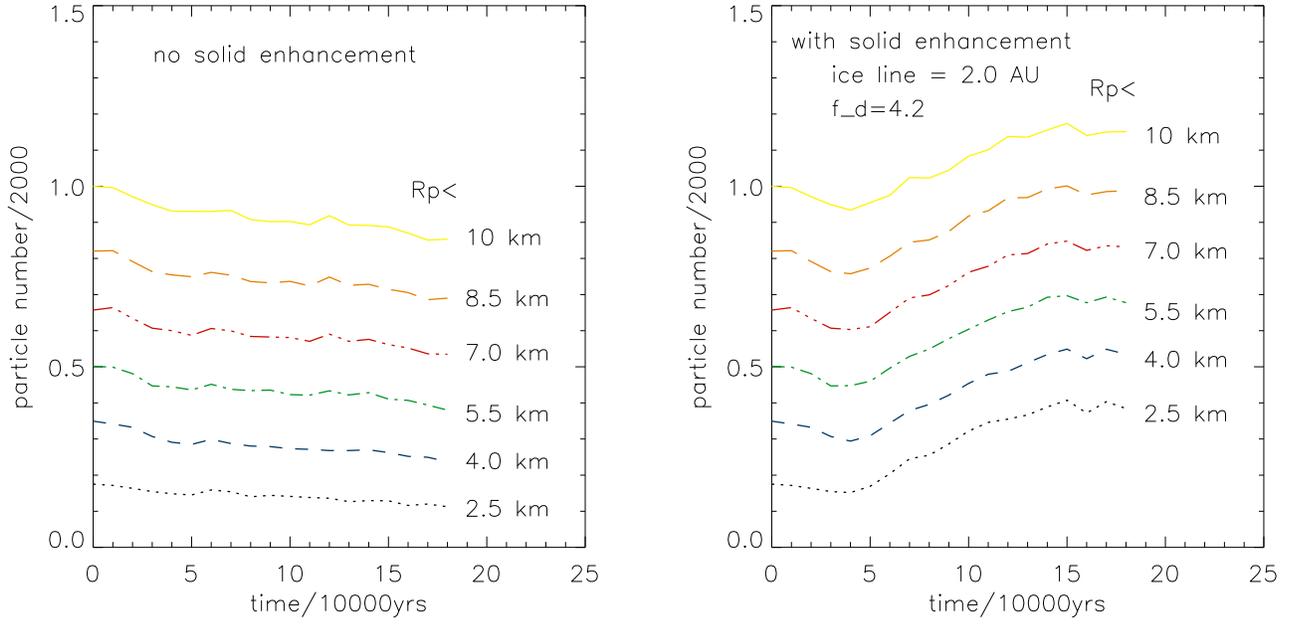}
  \caption{Number of Planetesimals within $a=1.4-1.6$ AU vs time for run 17 ($i_B=1^o$, $M_d=1$MMSN, including gas depletion). Left panel: case without considering the solid enhancement beyond ice line. Right panel: case including the solid enhancement of $f_g=4.2$ beyond ice line of 2.0 AU. }
   \end{center}
\end{figure}


\end{document}